\documentclass[english,aps,prper,reprint,showpacs,titlepage,longbibliography]{revtex4-1}   

\usepackage[T1]{fontenc}	
\usepackage[latin9]{inputenc}	
\usepackage{geometry}		
\usepackage{graphicx}
\usepackage[above,below]{placeins}	
\usepackage{times}

\usepackage{subcaption}

\usepackage{hyperref}  
\hypersetup{colorlinks=true,urlcolor=blue,citecolor=blue,linkcolor=blue}   
\usepackage{enumitem}          
\setlist{nosep}                 

\usepackage[]{natbib} 
\usepackage{hyperref}

\newcommand{\ngh}{\textcolor{black}}

\makeatletter
\renewcommand\frontmatter@abstractwidth{\dimexpr\textwidth-1in\relax}
\makeatother

\begin{document}

\begin{titlepage}

  \title{Preliminary evidence for available roles in mixed-gender and all-women lab groups}

  \author{N.G. Holmes}
    \author{Z. Yasemin Kalender}
  \affiliation{Laboratory of Atomic and Solid State Physics, Department of Physics, Cornell University, Ithaca, NY, 14850} 


  \begin{abstract}
  Group work during lab instruction can be a source of inequity between male and female students. In this preliminary study, we explored the activities male and female students take on during a lab session at a university in Denmark. Different from many studies, the class was majority-female, so three of the seven groups were all female and the rest were mixed-gender. We found that students in mixed-gender groups divide tasks in similar ways to mixed-gender groups at North American institutions, with men handling the equipment and women handling the computer more often. We also found that women in single-gender groups took on each of the available roles with approximately equal frequency, but women in single-gender groups spent more time on the equipment than students in mixed-gender groups. We interpret the results through poststructual gender theory and the notion of `doing physics' and `doing gender' in physics labs.\clearpage
  \end{abstract}

  \maketitle
\end{titlepage}

\clearpage

\section{Introduction and motivation}

Collaborative group work is pervasive in modern physics education. Laboratory (lab) instruction typically employs group work, where group members must coordinate a diverse array of tasks to accomplish a common goal. Previous work has found that male and female students (on average) divide such tasks inequitably~\cite{Doucette2020, Day2016, Quinn2020, Danielsson2009, Jovanovic1998}. In some cases, male students predominantly take on equipment manipulation~\cite{Danielsson2009, Jovanovic1998, Quinn2020} or data analysis~\cite{Day2016}, while female students predominantly take on managerial roles~\cite{Doucette2020, Danielsson2009} or computer work~\cite{Quinn2020}. Recent evidence suggests that men take on different roles depending on whether they are in single- or mixed-gender groups~\cite{Quinn2020}. 

Researchers are exploring multiple avenues to understand the nature of or mechanisms for these differential roles. In this work, we are inspired by the theoretical framework, developed by Danielsson and colleagues, that explores students `doing gender' and `doing physics' in the context of laboratory work~\cite{Danielsson2009, Danielsson2012}. This framework builds from post-structural gender theory in that gender is ``created and negotiated by the individual in response to a specific social setting''~\cite[][p.27]{Danielsson2012} and it is \textit{performed} as individuals take on positions available to them~\cite{Honan2000}. In physics lab activities, students can take on multiple positions or roles, such as handling the equipment, analyzing data, or documenting the activities. Because students typically work in small groups to complete their experiments, these tasks understandably get divided between the group members. Thus, as students negotiate `doing physics' during their investigations, the social setting requires them to also negotiate `doing gender'. 

With this lens, we interpret the emergence of distinct roles for male and female students in the previous literature as students' enacting of gender in the laboratory. For example, men handling the equipment more~\cite{Quinn2020, Jovanovic1998} may relate to performing masculinities such as tinkering~\cite{Danielsson2014, Gonsalves2016, Danielsson2012, Doucette2020, Danielsson2009}. Women handling the computer~\cite{Quinn2020} or engaging in other non-technical tasks~\cite{Day2016} may relate to performing feminities such as communication, secretary, or manager roles~\cite{Doucette2020, Danielsson2014, Gonsalves2016}.

While many of the observations of gendered division of tasks in physics labs took place in North America, the theoretical framework was developed by European researchers. Case studies across the United States, Canada, and Sweden, however, found remarkable similarities between the emergence of gender in doing physics and lab work~\cite{Gonsalves2016}. We sought to extend this international focus to evaluate whether the gendered division of tasks previously observed at North American institutions~\cite{Doucette2020, Day2016, Quinn2020} would also emerge at a European institution. 

Furthermore, the theoretical framework related to `doing gender' and `doing physics' was linked to associations between physics and masculinity, with women as visible, numerical minorities in the classroom or discipline~\cite{Gonsalves2016, Danielsson2009}. Similarly, the observations of gendered divisions of tasks took place in majority-male classrooms~\cite{Doucette2020, Day2016, Quinn2020}. Relatively little work has evaluated task divisions 
in female-majority classrooms and in all-women groups.

To pursue these questions, we conducted a preliminary, observational study of students at an institution in Denmark during a single lab session where women were the numerical majority. 
We aimed to shed light on two questions:
\begin{enumerate}
    \item Do male and female students in mixed-gender groups in Denmark exhibit similar inequitable divisions of tasks as those observed at North American institutions?
    \item In a female-majority classroom, how do students divide tasks in all-women versus mixed-gender groups?
\end{enumerate}
While the gender dynamics in Denmark are socially and politically different than in the United States, the work by Danielsson with students in Sweden~\cite{Danielsson2009, Danielsson2014, Danielsson2012, Gonsalves2016} led us to predict that the observations of gendered tasks may be similar to those in North America. We had no predictions for the behavior of women in single-gender groups, because the studies exploring student roles in labs described above had observed majority-male classrooms and focused on mixed-gender groups. 


\vspace{-0.5em}
\section{Methods}

In this preliminary study, we measured coarse-grain behaviors of students in an introductory mechanics course at a technical university in Denmark (majority engineering majors in the course and the institution). 

\vspace{-0.5em}
\subsection{Participants and instructional context}

The data were collected from observations of a single lab session. The observed lab session was the first and only lab session for the course. The course traditionally did not involve any lab activities, but the instructors were in the process of developing a new laboratory curriculum to accompany the course. The new lab curriculum aims to develop students' experimentation and critical thinking skills, while also supporting the conceptual physics relevant to the lecture. The observed lab session was a pilot session for the new lab curriculum and students volunteered to participate (and were given pizza as incentive).
The two-hour lab session was about projectile motion. Students were tasked with collecting data to determine the launcher's initial velocity and then to use that information, combined with concepts and equations from lecture, to hit an arbitrary target set by the instructors. 

Of 50 students enrolled in the course, 21 students volunteered to participate in the optional lab session. Although the course enrolled approximately equal numbers of men and women, the students in the lab session were predominantly female (16 women and 5 men). All students were physical science or engineering majors from a variety of sub-disciplines. 
This was their first college-level physics course, though most students completed high school-level physics. Students self-selected into groups of two to four students. Three instructors and two graduate teaching assistants attended the lab session to facilitate and observe. Two researchers were also present to observe and collect data, but did not interact with the students pedagogically. 

\vspace{-0.5em}
\subsection{Data collection}

The two researchers stood at the back of the lab room for the duration of the session. One of the observers took qualitative field notes of the students' activities. All instruction and most conversations took place in Danish, and neither researcher spoke Danish. Instructors periodically checked in with the researchers to provide context for the conversations that could not be interpreted based on the visual activities. 


The second researcher took structured observations of student activities based on a protocol used in Refs.~\cite{Quinn2020, Day2016}. The protocol has demonstrated strong inter-rater reliability~\cite{Quinn2020} and the observer in this study had previously used this protocol with sufficient interrater reliability (citation omitted for confidential review). With this protocol, the researcher recorded, in two to three minute intervals, 
whether each student had their hands on the experimental equipment, a computer or laptop, or pencil and paper, or whether they were talking to their group, another group, or the instructor. Any other behaviors were coded as `Other.' \ngh{These activities have been previously compared to qualitative descriptions of the students' experimentation tasks, such that we can approximately infer experimentation roles from their tasks~\cite{Quinn2020}. That is, handling the equipment reflects either setting up (or taking down) the apparatus or collecting data, handling the computer reflects entering data, analyzing the data, or documenting methods in lab notes, handling paper reflects making notes or consulting the lab manual, and so on.} The data indicated that very few students interacted with other group members, so this code was collapsed with `Other.' If the student was shifting activities, the observer waited no more than 10 seconds to identify the dominant activity. With 21 students present, each student's behavior could be documented within the two to three minute window, moving sequentially around the room. A total of 43 intervals were coded for each student.




\vspace{-0.5em}
\subsection{Analysis}

The coding protocol provides coarse-grain, efficient, and broad information about all students in the class. The protocol does not attend to fine-grain interactions or student conversation. From previous work with the protocol, the coding categories effectively estimate and distinguish forms of student engagement in various lab activities (e.g., analyzing data is reserved to handling the computer, considering the experimental set up or collecting data are reserved to handling the equipment, taking notes or consulting the lab manual are reserved to handling paper)~\cite{Quinn2020}. The coarseness of the data places limitations on the interpretations, but allows a broad sample of all students over a long period of time, providing an estimate of students' roles and activities in the lab.

We calculated the total number of coded intervals for each student for each activity. 
We then produced scatter plots of the frequencies of intervals in which students were coded as engaging with each type of activity. We also explored the total number of coded intervals by each student for each activity, comparing within and between groups. 
\ngh{We summarize these data into two figures. The first looks across all groups to identify overall differences between men and women in mixed- and single-gender groups, including the mean and standard error of the number of coded intervals across demographics. The second looks within each group to compare the behaviors of each group member compared to their peers.} 
Given the small sample size, statistical comparisons would not be meaningful, so we rely only on the graphical analyses. 

Based on previous cluster analyses~\cite{Quinn2020}, we used the total frequencies across the lab period to estimate the students' \ngh{dominant roles by evaluating each student's engagement in the activity compared to the rest of the class. That is, we converted} 
each student's total frequency for each activity to a $z$-score relative to the rest of the class. For each person for each activity, a $z$-score is calculated as:
 \begin{equation}
     z = \frac{F_i - \overline{F}}{\sigma}
 \end{equation}
 where $F_i$ is the total frequency for the individual for the given activity, and $\overline{F}$ and $\sigma$ are the average and standard deviation of the class' total frequency for the given activity, respectively. The $z$-score, therefore, indicates by how many standard deviations the individual's frequency differs from the class average. A $z$-score greater than zero indicates that the student performed the activity more than average, while a $z$-score less than zero indicates that the student performed the activity less than average. A student's dominant activity was identified as the activity with their largest $z$-score compared to the rest of the class. This method takes into account that, on average, students spent unequal amounts of time on different tasks. 
 We then compared the proportion of students with each type of dominant activity by group type and gender.

\begin{figure*}[hbt!]
    \centering
    \includegraphics[width=0.9\linewidth]{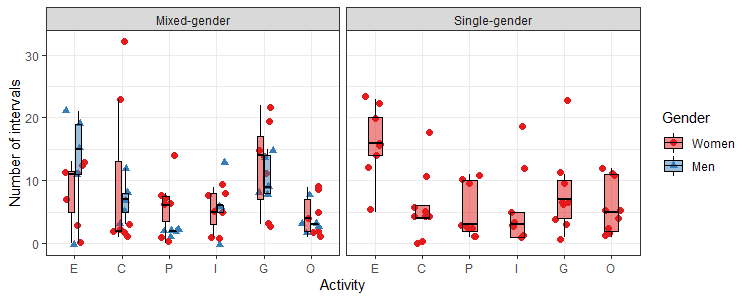}
    \caption{\ngh{Scatter and box plot} of the number of coded intervals for each activity \ngh{(\textbf{E}quipment, \textbf{C}omputer, \textbf{P}aper, \textbf{I}nstructor, \textbf{G}roup, and \textbf{O}ther)} for each student by gender and group type. \ngh{The box plots give the median (thick central line) and the first and third quartiles, and the whiskers extend across 1.5 times the interquartile range. The box plots for men on paper, discussing with the instructor, and doing other are too small to be visible.}
    } \vspace{-1em}
    \label{fig:histogram}
\end{figure*}

\vspace{-1em}
\section{Results}


Figure~\ref{fig:histogram} shows the distribution of the number of students coded as engaging in each type of activity for various numbers of intervals, \ngh{broken out by gender and group type (mixed- versus single-gender groups)}. 
Within mixed-gender groups, we see some evidence of men and women taking on different tasks. For example, \ngh{on average, men spent} more time on the equipment than women and women spent more time on paper than men. Women's use of the computer was \ngh{highly variable,} 
with some women almost never using the computer and some almost exclusively using the computer. Discussing with the instructor, discussing with their group, and participating in Other activities were similar between men and women in mixed-gender groups. These results appear similar to those observed previously~\cite{Quinn2020}. 

\ngh{Comparing the two panels, we can identify some interesting contrasts between students in mixed- and single-gender groups.} 
The figure shows that women in single-gender groups spent more time on the equipment than women in mixed-gender groups \ngh{and, on average, spend more time on the equipment than they do on any of the other tasks}. \ngh{Three students in mixed-gender groups seemed to spend almost no time on the equipment, while no students in single-gender groups spent that little time on the equipment. 
Time spent on the computer is much less variable than for women in mixed-gender groups and the average is again more comparable to the use by men. Other than one student, women in single-gender groups seemed to spend much less time discussing with their group than students in mixed-gender group.} Time spent on paper or doing Other is quite similar between \ngh{students} in the two types of groups. 

\begin{figure*}
    \centering
    \includegraphics[width=\linewidth]{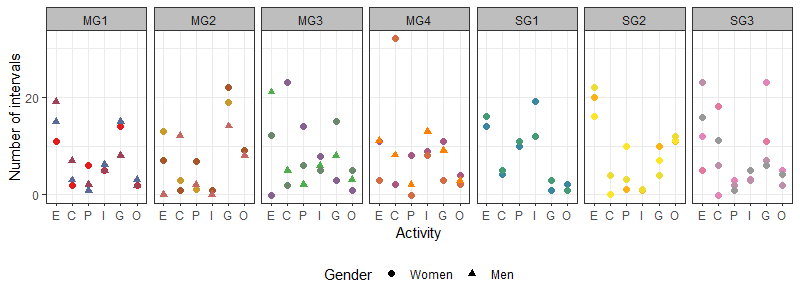}\vspace{-1em}
    \caption{Scatter plot of the number of intervals with which each student (shown as a dot) was coded as engaging in each type of behavior. The \textit{jitter} function was applied so that individual dots did not sit on top of each other. Dots close together mean that all group members spent similar amounts of time doing an activity, while spread suggests tasks were divided. Groups labeled `SG' represent single-gender groups and those labeled `MG' represented mixed-gender groups.}
    \label{fig:scatter}\vspace{-1em}
\end{figure*}

Figure~\ref{fig:scatter} provides context for each individual's behavior relative to their group members. The figure presents the number of intervals for which each student was coded engaging in each activity. \ngh{We have colored each student so that they can be identified for each task.} We use this representation \ngh{predominantly} to evaluate whether students shared tasks or divided tasks. \ngh{We identify groups} that shared tasks (MG1, MG2, SG1, and SG2) \ngh{as ones whose points are generally clustered together. That is,} all group members performed an activity for similar numbers of intervals. 
\ngh{We identify groups} that divided tasks (SG3 and MG3) \ngh{as ones whose points are quite spread out. That is,} group members performed an activity for a different number of intervals, with some students particularly high and others particularly low. MG4 generally share tasks, except for one female student who almost exclusively uses the computer \ngh{and does little else}. Overall, sharing or dividing tasks does not seem to depend on group composition (i.e., single- versus mixed-gender groups). 



\begin{figure}[htbp]
    \centering
    \includegraphics[width=\linewidth]{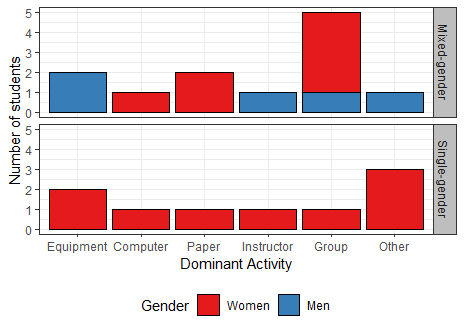}
    \caption{Histogram of the number of students identified as dominant in each type of activity relative to the class, broken down by gender and group type.}
    \label{fig:role}
\end{figure}

The results are further illuminated by the representation in Figure~\ref{fig:role}, which shows the number of students that were identified as dominant in each type of activity, broken down by gender and group type.
A student's dominant role was identified as the activity the student performs most above average, compared to the rest of the class. We see that women in single-gender groups were represented across all activity roles, while women in mixed-gender groups were dominant only on computer, paper, or discussing with their group. Men were never dominant computer or paper users. 


\vspace{-0.5em}
\section{Discussion}

We conducted a preliminary, observational study to understand how male and female students at an institution in Denmark divide tasks during a lab. 
In line with existing literature~\cite{Gonsalves2016}, the gender patterns of task allocation were similar to results found in North America~\cite{Quinn2020, Doucette2020}. That is, we find that, in mixed-gender groups, men spent more time handling equipment than women, and that women were dominantly engaged in working on the computer, discussing with the group, or handling paper worksheets. 

We uniquely find that women in single-gender groups \ngh{spent more time handling the equipment than students in mixed-gender groups. Furthermore, women in single-gender groups were equally likely to dominate any type of activity, with dominant tasks being defined relative to the whole class.} 
In the mixed-gender groups, in contrast, men and women \ngh{each} had roles not available to them. For example, no women in mixed-gender groups were dominant on the equipment, \ngh{talking to the instructor, or doing other tasks} and no men were dominant computer or paper users. \ngh{We interpret this result through our theoretical framework: in mixed-gender groups, students do not take on particular roles because those roles are not available them.} The lack of available roles for men and women in mixed-gender groups supports the \ngh{notion} that students were navigating \emph{both} `doing physics' and `doing gender'~\cite{Danielsson2009, Danielsson2012}. \ngh{That is, while all of these roles are necessary for doing the experiment, some are perceived as more available to women than men and vice versa. Equipment-handling in particular is said to be ``a doing of a particular classed masculinity''~\cite[][p.488]{Danielsson2014}. Our evidence supports this claim in that no women in mixed-gender groups were dominant equipment users. However, our data also suggest that these tasks are not inherently gendered. If we ignore gender in Fig.~\ref{fig:role}, students were similarly dominant across the activities in each group type. As a reminder, dominant activities are defined relative to \emph{all} students in the class, not within groups.} 
\ngh{We infer, with support from the framework, that} all tasks were available to \ngh{the women in single gender groups}~\cite{Honan2000} and that \ngh{these} students were not negotiating `doing gender'~\cite{Danielsson2009, Danielsson2012}, but were simply negotiating `doing physics' \ngh{as they each take up different tasks}. 


Perhaps surprisingly, however, we do not see that \ngh{students in either group type} 
necessarily divide the work equally. That is, some groups seemed to divide-and-conquer 
\ngh{while others} seemed to share the work equally, \ngh{with no patterns between} 
single- and mixed-gender groups. Thus, women in single-gender groups are still apt to take on roles where one or two students dominate, for example, the equipment (e.g., group SG3). This raises questions about whether group composition (based on gender) predicts the equity of the group, as suggested by previous literature~\cite{Heller1992, Tanner2013, Dasgupta2015}.  

Because the data are limited in multiple ways, all results should be considered tentative. We highlight, in particular, that a higher proportion of female students attended the lab than were enrolled in the course, which suggests that the students may not be representative of the general course population. However, the population provided a unique sampling opportunity to quickly and roughly probe hypotheses from existing literature, and thus warrant future study. In addition to collecting more data, future work should also explore how students navigate into these roles. While previous work found that most task allocations were not overt~\cite{Quinn2020}, 
analysis of student positioning and body language in the labs may shed light on how roles are being assigned or assumed. 

\acknowledgments{This work was supported by a Denmark Technical University Grant for the Development of Teaching Quality and through the Cornell University Active Learning Intitiative. We would like to sincerely thank Kristoffer Haldrup, Carsten Knudsen, and Ole Trinhammer for their leadership on the lab practicals at DTU.}

\newpage

\bibliographystyle{apsrev} 
\bibliography{bib.bib} 

\begin{thebibliography}{12}
\expandafter\ifx\csname natexlab\endcsname\relax\def\natexlab#1{#1}\fi
\expandafter\ifx\csname bibnamefont\endcsname\relax
  \def\bibnamefont#1{#1}\fi
\expandafter\ifx\csname bibfnamefont\endcsname\relax
  \def\bibfnamefont#1{#1}\fi
\expandafter\ifx\csname citenamefont\endcsname\relax
  \def\citenamefont#1{#1}\fi
\expandafter\ifx\csname url\endcsname\relax
  \def\url#1{\texttt{#1}}\fi
\expandafter\ifx\csname urlprefix\endcsname\relax\def\urlprefix{URL }\fi
\providecommand{\bibinfo}[2]{#2}
\providecommand{\eprint}[2][]{\url{#2}}

\bibitem[{\citenamefont{Doucette et~al.}(2020)\citenamefont{Doucette, Clark,
  and Singh}}]{Doucette2020}
\bibinfo{author}{\bibfnamefont{D.}~\bibnamefont{Doucette}},
  \bibinfo{author}{\bibfnamefont{R.}~\bibnamefont{Clark}}, \bibnamefont{and}
  \bibinfo{author}{\bibfnamefont{C.}~\bibnamefont{Singh}},
  \bibinfo{journal}{European Journal of Physics} \textbf{\bibinfo{volume}{41}},
  \bibinfo{pages}{035702} (\bibinfo{year}{2020}).

\bibitem[{\citenamefont{Day et~al.}(2016)\citenamefont{Day, Stang, Holmes,
  Khumar, and Bonn}}]{Day2016}
\bibinfo{author}{\bibfnamefont{J.}~\bibnamefont{Day}},
  \bibinfo{author}{\bibfnamefont{J.~B.} \bibnamefont{Stang}},
  \bibinfo{author}{\bibfnamefont{N.~G.} \bibnamefont{Holmes}},
  \bibinfo{author}{\bibfnamefont{D.}~\bibnamefont{Khumar}}, \bibnamefont{and}
  \bibinfo{author}{\bibfnamefont{D.~A.} \bibnamefont{Bonn}},
  \bibinfo{journal}{Physical Review Physics Education Research}
  \textbf{\bibinfo{volume}{12}}, \bibinfo{pages}{020104}
  (\bibinfo{year}{2016}).

\bibitem[{\citenamefont{Quinn et~al.}(2020)\citenamefont{Quinn, Kelley, McGill,
  Smith, Whipps, and Holmes}}]{Quinn2020}
\bibinfo{author}{\bibfnamefont{K.~N.} \bibnamefont{Quinn}},
  \bibinfo{author}{\bibfnamefont{M.~M.} \bibnamefont{Kelley}},
  \bibinfo{author}{\bibfnamefont{K.~L.} \bibnamefont{McGill}},
  \bibinfo{author}{\bibfnamefont{E.~M.} \bibnamefont{Smith}},
  \bibinfo{author}{\bibfnamefont{Z.}~\bibnamefont{Whipps}}, \bibnamefont{and}
  \bibinfo{author}{\bibfnamefont{N.}~\bibnamefont{Holmes}},
  \bibinfo{journal}{Physical Review Physics Education Research}
  \textbf{\bibinfo{volume}{16}}, \bibinfo{pages}{010129}
  (\bibinfo{year}{2020}), ISSN \bibinfo{issn}{2469-9896},
  \urlprefix\url{https://link.aps.org/doi/10.1103/PhysRevPhysEducRes.16.010129}.

\bibitem[{\citenamefont{Danielsson and Linder}(2009)}]{Danielsson2009}
\bibinfo{author}{\bibfnamefont{A.~T.} \bibnamefont{Danielsson}}
  \bibnamefont{and} \bibinfo{author}{\bibfnamefont{C.}~\bibnamefont{Linder}},
  \bibinfo{journal}{Gender and Education} \textbf{\bibinfo{volume}{21}},
  \bibinfo{pages}{129} (\bibinfo{year}{2009}).

\bibitem[{\citenamefont{Jovanovic and King}(1998)}]{Jovanovic1998}
\bibinfo{author}{\bibfnamefont{J.}~\bibnamefont{Jovanovic}} \bibnamefont{and}
  \bibinfo{author}{\bibfnamefont{S.~S.} \bibnamefont{King}},
  \bibinfo{journal}{American Educational Research Journal}
  \textbf{\bibinfo{volume}{35}}, \bibinfo{pages}{477} (\bibinfo{year}{1998}).

\bibitem[{\citenamefont{Danielsson}(2012)}]{Danielsson2012}
\bibinfo{author}{\bibfnamefont{A.~T.} \bibnamefont{Danielsson}},
  \bibinfo{journal}{Gender and Education} \textbf{\bibinfo{volume}{24}},
  \bibinfo{pages}{25} (\bibinfo{year}{2012}).

\bibitem[{\citenamefont{Honan et~al.}(2000)\citenamefont{Honan, Knobel, Baker,
  and Davies}}]{Honan2000}
\bibinfo{author}{\bibfnamefont{E.}~\bibnamefont{Honan}},
  \bibinfo{author}{\bibfnamefont{M.}~\bibnamefont{Knobel}},
  \bibinfo{author}{\bibfnamefont{C.}~\bibnamefont{Baker}}, \bibnamefont{and}
  \bibinfo{author}{\bibfnamefont{B.}~\bibnamefont{Davies}},
  \bibinfo{journal}{Qualitative Inquiry} \textbf{\bibinfo{volume}{6}},
  \bibinfo{pages}{9} (\bibinfo{year}{2000}).

\bibitem[{\citenamefont{Danielsson}(2014)}]{Danielsson2014}
\bibinfo{author}{\bibfnamefont{A.~T.} \bibnamefont{Danielsson}},
  \bibinfo{journal}{Cultural Studies of Science Education}
  \textbf{\bibinfo{volume}{9}}, \bibinfo{pages}{477} (\bibinfo{year}{2014}).

\bibitem[{\citenamefont{Gonsalves et~al.}(2016)\citenamefont{Gonsalves,
  Danielsson, and Pattersson}}]{Gonsalves2016}
\bibinfo{author}{\bibfnamefont{A.~J.} \bibnamefont{Gonsalves}},
  \bibinfo{author}{\bibfnamefont{A.~T.} \bibnamefont{Danielsson}},
  \bibnamefont{and}
  \bibinfo{author}{\bibfnamefont{H.}~\bibnamefont{Pattersson}},
  \bibinfo{journal}{Physical Review Physics Education Research}
  \textbf{\bibinfo{volume}{12}}, \bibinfo{pages}{020120}
  (\bibinfo{year}{2016}).

\bibitem[{\citenamefont{Heller and Hollabaugh}(1992)}]{Heller1992}
\bibinfo{author}{\bibfnamefont{P.}~\bibnamefont{Heller}} \bibnamefont{and}
  \bibinfo{author}{\bibfnamefont{M.}~\bibnamefont{Hollabaugh}},
  \bibinfo{journal}{American Journal of Physics} \textbf{\bibinfo{volume}{60}},
  \bibinfo{pages}{627} (\bibinfo{year}{1992}), ISSN \bibinfo{issn}{00029505},
  \urlprefix\url{http://link.aip.org/link/?AJP/60/627/1{\&}Agg=doi
  http://ajp.aapt.org/resource/1/ajpias/v60/i7/p627{\_}s1}.

\bibitem[{\citenamefont{Tanner}(2013)}]{Tanner2013}
\bibinfo{author}{\bibfnamefont{K.~D.} \bibnamefont{Tanner}},
  \bibinfo{journal}{Cell Biology Education} \textbf{\bibinfo{volume}{12}},
  \bibinfo{pages}{322} (\bibinfo{year}{2013}), ISSN \bibinfo{issn}{1931-7913},
  \urlprefix\url{http://www.lifescied.org/cgi/doi/10.1187/cbe.13-06-0115}.

\bibitem[{\citenamefont{Dasgupta et~al.}(2015)\citenamefont{Dasgupta, Scircle,
  and Hunsinger}}]{Dasgupta2015}
\bibinfo{author}{\bibfnamefont{N.}~\bibnamefont{Dasgupta}},
  \bibinfo{author}{\bibfnamefont{M.~M.} \bibnamefont{Scircle}},
  \bibnamefont{and}
  \bibinfo{author}{\bibfnamefont{M.}~\bibnamefont{Hunsinger}},
  \bibinfo{journal}{Proceedings of the National Academy of Sciences of the
  United States of America} \textbf{\bibinfo{volume}{112}},
  \bibinfo{pages}{4988} (\bibinfo{year}{2015}), ISSN \bibinfo{issn}{1091-6490},
  \urlprefix\url{http://www.ncbi.nlm.nih.gov/pubmed/25848061
  http://www.pubmedcentral.nih.gov/articlerender.fcgi?artid=PMC4413283}.

\end{thebibliography}

\end{document}